\documentclass{article}
\usepackage{amssymb}
\usepackage{amsfonts}
\usepackage{color}
\usepackage{graphicx}
\usepackage{epsfig}
\usepackage{theorem}
\newtheorem{theorem}{Theorem}

\newtheorem{definition}[theorem]{Definition}

\newtheorem{proposition}[theorem]{Proposition}
{\theorembodyfont{\rm}
\newtheorem{example}[theorem]{Example}

}
\newenvironment{proof}[1][Proof]{\noindent\textbf{#1.\ }}
{\hfill \rule{1ex}{1ex}}

\newcommand{\sC}{{\mathbb C}}

\newcommand{\sN}{{\mathbb N}}

\newcommand{\sR}{{\mathbb R}}

\newcommand{\sT}{{\mathbb T}}

\newcommand{\sZ}{{\mathbb Z}}
\newcommand{\cA}{{\mathcal A}}

\newcommand{\cC}{{\mathcal C}}

\newcommand{\cH}{{\mathcal H}}

\newcommand{\cK}{{\mathcal K}}
\newcommand{\cL}{{\mathcal L}}
\newcommand{\cM}{{\mathcal M}}

\newcommand{\cW}{{\mathcal W}}
\newcommand{\cX}{{\mathcal X}}

\newcommand{\diag}{\mbox{\rm diag} \,}
\newcommand{\op}{{\rm op} \,}

\newcommand{\opp}{{\rm op}_p \,}

\newcommand{\spec}{{\rm sp} \,}
\newcommand{\spd}{{\rm sp}_{dis} \,}
\newcommand{\spe}{{\rm sp}_{ess} \,}
\newcommand{\qed}{\rule{1ex}{1ex}}
\begin{document}

\title{Essential spectra of difference operators on $\sZ^n$-periodic 
graphs}
\author{V. S. Rabinovich, S. Roch \\
Vladimir S. Rabinovich, Instituto Polit\'{e}cnico Nacional, \\
ESIME-Zacatenco, Av. IPN, edif.1, M\'{e}xico D.F., \\
07738, M\'{E}XICO, \ e-mail: vladimir.rabinovich@gmail.com \\
Steffen Roch, Technische Universit\"{a}t Darmstadt, \\
Schlossgartenstrasse 7, 64289 Darmstadt, Germany, \\
e-mail: roch@mathematik.tu-darmstadt.de}
\date{}
\maketitle
\begin{abstract}
Let $(\cX, \, \rho)$ be a discrete metric space. We suppose that 
the group $\sZ^n$ acts freely on $X$ and that the number of orbits
of $X$ with respect to this action is finite. Then we call $X$ a 
$\sZ^n$-periodic discrete metric space. We examine the Fredholm 
property and essential spectra of band-dominated operators on 
$l^p(X)$ where $X$ is a $\sZ^n$-periodic discrete metric space. 
Our approach is based on the theory of band-dominated operators 
on $\sZ^n$ and their limit operators.

In case $X$ is the set of vertices of a combinatorial graph, the 
graph structure defines a Schr\"{o}dinger operator on $l^p(X)$ in 
a natural way. We illustrate our approach by determining the essential 
spectra of Schr\"{o}dinger operators with slowly oscillating potential 
both on zig-zag and on hexagonal graphs, the latter being related to 
nano-structures. 
\end{abstract}
\section{Introduction} \label{sect1}
In the last years, spectral properties of Schr\"{o}dinger operators 
on quantum graphs have attracted a lot of attention due to their 
interesting mathematical properties and due to existing and expected 
applications in nano-structures as well (see, for instance, 
\cite{Berkol,Harris,Saito}). Quantum graph models also occur in
chemistry \cite{Paul,RuSch} and physics \cite{Berkol,Kuch1} (see also
the references therein). The spectral properties of Schr\"{o}dinger
operators on quantum graphs considered by P. Kuchment and
collaborators in a series of papers
\cite{KuchSp,Kuch1,Kuch2,Kuch3,KuchPost,KuchVainberg}. Direct and
inverse spectral problems for Schr\"{o}dinger operators on graphs
connected with zig-zag carbon nano-tubes was considered in
\cite{Kor1,Kor2}.

It was shown in \cite{Kuch2,Kuch3} that the spectral analysis of
quantum Hamiltonian on periodic graphs splits into two parts: the
spectral analysis of a Hamiltonian on a single edge, and the
spectral analysis on a combinatorial graph. This observation makes 
difference operators on combinatorial graphs to an essential tool
in the theory of differential operators on quantum graphs. 

The main theme of this paper is the essential spectrum of 
difference operators (with the Schr\"{o}dinger operators as a 
prominent example) acting on the spaces $l^p(X)$ where $X$ is 
the set of the vertices of a combinatorial graph $\Gamma$. 
We exclusively consider discrete graphs $\Gamma$ on which the 
group $\sZ^n$ acts freely and which have a finite fundamental 
domain with respect to this action. 

We introduce a Banach algebra $\cA_p (X)$ of so-called
band-dominated difference operators $l^p(X)$ for $1 < p
< \infty$. Following \cite{RRS1,RRS2} and \cite{RRSB}, we
introduce for each operator $A \in \cA_p (X)$ a family 
$\opp (A)$ of {\em limit operators} of $A$, and we show that 
an operator $A \in \cA_p (X)$ is {\em Fredholm} on $l^p (X)$ 
if and only if all operators in $\opp (A)$ are invertible and 
if the norms of their inverses are uniformly bounded. In general, 
the limit operators of an operator $A$ are simpler objects than 
the operator $A$ itself. Thus, the limit operators method often 
provides an effective tool to study the Fredholmness of operators 
in $\cA_p (X)$. 

For operators in the so-called Wiener algebra $\cW(X)$ (which 
is a non-closed subalgebra of every algebra $\cA_p (X)$, the 
uniform boundedness of norms of inverse operators to limit 
operators follows already from their invertibility. This basic 
fact implies the useful identity 
\begin{equation} \label{1}
\spe  A = \bigcup_{A_h \in \op A} \spec A_h
\end{equation}
where the set of the limit operators of $A$, the spectra 
$\spec A_h$ of the limit operators of $A$ and, hence, also the 
essential spectrum $\spe A$ of $A$ are independent of $p$.

In case $X = \sZ^n$, formula (\ref{1}) was obtained in \cite{RRS1}, 
see also \cite{RRSB}. In \cite{RRJP}, we applied this formula to 
study electromagnetic Schr\"{o}dinger operators on the lattice 
$\sZ^n$. In particular, we determined the essential spectra of 
the Hamiltonian of the 3-particle problem on $\sZ^n$.

In \cite{RJMP}, one of the authors obtained an identity similar to 
(\ref{1}) for perturbed pseudodifferential operators on $\sR^n$.
He applied this result to study the location of the essential spectra 
of electromagnetic Schr\"{o}dinger operators, square-root
Klein-Gordon, and Dirac operators under general assumptions with
respect to the behavior of magnetic and electric potentials at
infinity. By means of this method, also a very simple and transparent
proof of the well known Hunziker, van Winter, Zjislin theorem
(HWZ-Theorem) on the location of essential spectra of multi-particle
Hamiltonians was obtained.

It should be noted that formulas similar to (\ref{1}) have been
obtained independently (but later) in \cite{LastSimon} by means 
of admissible geometric methods. We also mention the papers
\cite{Gorgescu1,Gorgescu2,Mantoiu,AMP} and the references therein 
where $C^*$-algebra techniques have been applied to study essential 
spectra of Schr\"{o}dinger operators.

The present paper is organized as follows. In Section 2 we collect 
some auxiliary material from \cite{RRSB} on matrix band-dominated
operators on the lattice $\sZ^n$. In Section 3 we introduce the
Banach algebra $\cA_p (X)$ of band-dominated operators acting
on $l^p (X)$ where $X$ is a periodic discrete metric space  
on which the group $\sZ^n$ acts freely. We construct an isomorphism
between the Banach algebra $\cA_p (X)$ and the Banach algebra
$\cA_p (\sZ^n, \, \sC^N)$ of all (block) band-dominated operators on
$l^p(\sZ^n, \,  \sC^N)$ where $N$ is the number of points in the
fundamental domain of $X$ with respect to the action of $\sZ^n$.
Applying this isomorphism and the results of Section 2, we derive
necessary and sufficient conditions for $A \in \cA_p(X)$ to be
a Fredholm operator. We also introduce a Wiener algebra $\cW (X)$ and 
derive formula (\ref{1}) for operators in $\cW (X)$.

In Section 4 we introduce the class of periodic band-dominated
operators. We say that $A \in \cA_p (X)$ is a periodic
operator if it commutes with each operator $L_h$ of left shift 
by $h \in \sZ^n$ on $l^p(X)$. Note that, for periodic operators,
$\spe A = \spec A$. With each periodic operator $A \in \cW (X)$,
we associate a continuous function $\sigma_A : \sT^n \to \sC^{N
\times N}$, called the symbol of $A$. In the terminology of
\cite{Kuch1,Kuch2}, $\sigma_A (t)$ is just the Floquet transform of 
$A$. We prefer to follow the theory of discrete convolutions and 
use the discrete Fourier transform to define $\sigma_A$.

Let $\lambda_j (t)$, $j = 1, \, \ldots, \, N$ be the eigenvalues of
$\sigma_A (t)$. Then
\[
\spec A = \bigcup_{j=1}^N \cC_j (A)
\]
where $\cC_j (A) := \{ \lambda \in \sC : \lambda = \lambda_j(t),
\; t \in \sT^n \}$. If $A$ is a self-adjoint operator on $l^2
(X)$, then the $\cC_j (A)$ can be identified with segments.

In Section 5 we consider operators in the Wiener algebra $\cW
(X)$ with slowly oscillating coefficients. These operators
are distinguished by two remarkable properties: their limit
operators are periodic operators, and all limit operators belong
to the Wiener algebra again. Via formula (\ref{1}) we thus obtain
a complete description of the essential spectra of operators with
slowly oscillating coefficients.

In Section 6 we apply these results to Schr\"{o}dinger operators
with slowly oscillating electrical potentials. As already mentioned, 
every $\sZ^n$-periodic graph induces a related Schr\"{o}dinger 
operator in a natural way (it is only this place where the graph 
structure becomes important). As illustrations we calculate the essential 
spectra of Schr\"{o}dinger operators with slowly oscillating potentials 
on the {\em zig-zag} graph and on the {\em hexagonal} graph. Some 
other spectral problems on such graphs which are connected with carbon 
nano-structures were considered in \cite{Kor1,Kor2,KuchPost}.

In Section 7 we examine the essential spectrum of the Hamiltonian of
the motion of two particles on a periodic graph $\Gamma$ around a
heavy nucleus. For the lattice $\Gamma = \sZ^n$ we considered this
problem in \cite{RRJP}. See also the papers
\cite{AlbaverioLakaev,AlbLakMum,LakaevMuminov,Mattis,Mogilner} and
the references therein which are devoted to discrete multi-particle
problems.

The limit operators approach does also apply to study the
essential spectrum of pseudodifferential operators on periodic
quantum graphs. We plan to develop these ideas in a forthcoming
paper.

The authors are grateful for the support by CONACYT (Project 43432)
and by the German Research Foundation (Grant 444 MEX-112/2/05). 
\section{Band-dominated operators on $\sZ^n$} \label{sect2}
In this section we fix some notations and recall some facts
concerning the Fredholm property of band-dominated operators on
$l^p(\sZ^n)$. The Fredholm properties of these operators are
fairly well understood. All details can be found in \cite{RRS1};
see also the monograph \cite{RRSB} for a comprehensive account.

We will use the following notations. Given a Banach space $X$, let
$\cL(X)$ refer to the Banach algebra of all bounded linear operators
on $X$ and $\cK(X)$ to the closed ideal of the compact operators.
An operator $A \in \cL(X)$ is called a {\em Fredholm operator} if
its kernel $\ker A := \{ x \in X : Ax = 0 \}$ and its cokernel
$\mbox{coker} \, A := X / A(X)$ are finite dimensional linear
spaces. Equivalently, $A$ is Fredholm if the coset $A + \cK(X)$ is
invertible in the Calkin algebra $\cL(X)/\cK(X)$. The {\em essential
spectrum} of $A$ is the set of all complex numbers $\lambda$ for
which the operator $A - \lambda I$ is not Fredholm on $X$, whereas
the discrete spectrum of $A$ consists of all isolated eigenvalues of
finite multiplicity. We denote the essential spectrum of $A$ by
$\spe A$, the discrete spectrum by $\spd A$, and the usual spectrum
by $\spec A$. Sometimes we also write $\spec (A : X \to X)$ instead 
of $\spec A$ in order to emphasize the underlying space $X$ (with 
obvious modifications for the essential and the discrete spectrum). 
Clearly,
\[
\spd (A) \subseteq \spec (A) \setminus \spe (A)
\]
for every operator $A \in \cL(X)$. If $A$ is a self-adjoint
operator, then equality holds in this inclusion.

Let $p \ge 1$ be a real number and $n$ a positive integer. As usual,
we write $l^p (\sZ^n)$ for the Banach space of all functions $u :
\sZ^n \to \sC$ for which
\[
\|u\|_{l^p(\sZ^n)}^p := \sum_{x \in \sZ^n} |u(x)|^p < \infty
\]
and $l^\infty (\sZ^N)$ for the Banach space of all bounded functions 
$u : \sZ^n \to \sC$ with norm
\[
\|u\|_{l^\infty (\sZ^n)} := \sup_{x \in \sZ^n} |u(x)|.
\]
For every positive integer $N$, let $l^p(\sZ^n)^N$ stand for the
Banach space of all vectors $u = (u_1, \, \ldots, \, u_N)$ of   
functions $u_i \in l^p(\sZ^n)$ with norm
\[
\|u\|_{l^p (\sZ^n)^N}^p := \sum_{i = 1}^N \|u_i\|^p_{l^p (\sZ^n)}
\]
Likewise, one can identify $l^p(\sZ)^N$ with the Banach space 
$l^p(\sZ^n, \, \sC^N)$ of all functions $u : \sZ^n \to \sC^N$ for 
which
\[
\|u\|_{l^p(\sZ^n, \, \sC^N)}^p := \sum_{x \in \sZ^n} \sum_{i=1}^N 
|u_j(x)|^p < \infty.
\]
Clearly, the Banach spaces $l^p (\sZ^n)^N$ and $l^p(\sZ^n, \, 
\sC^N)$ are isometric to each other. We also consider the Banach 
spaces $l^\infty (\sZ^n)^N$ and $l^\infty (\sZ^n, \, \sC^N)$ with 
norms   
\[
\|u\|_{l^\infty (\sZ^n)^N} := \sup_{1 \le i \le N}
\|u_i\|_{l^\infty(\sZ^n)} 
\]
and
\[ 
\|u\|_{l^\infty (\sZ^n, \, \sC^N)} := \sup_{x \in \sZ^n} 
\sup_{1 \le i \le N} |u_i(x)|.
\]
Again, these spaces are isometric to each other in a natural way.
Note also that $l^\infty (\sZ^n, \, \sC^{N \times N})$ can be made 
to a $C^*$-algebra by providing the matrix algebra $\sC^{N \times N}$
with a $C^*$-norm.

We consider operators on $l^p (\sZ^n, \sC^N)$ which are constituted 
by shift operators and by operators of multiplication by bounded
functions. The latter are defined as follows: For $\alpha \in
\sZ^n$, the shift operator $V_\alpha$ is the isometry acting on
$l^p(\sZ^n, \sC^N)$ by $(V_\alpha u)(x) := u(x - \alpha)$. Further, 
each function $a$ in $l^\infty (\sZ^n, \, \sC^{N \times N})$ induces 
a multiplication operator $aI$ on $l^p(\sZ^n, \sC^N)$ via $(au)(x) 
:= a(x) u(x)$. Clearly,
\[
\|aI\|_{\cL(l^p(\sZ^n, \, \sC^N))} = \|a\|_{l^\infty (\sZ^n, \, 
\sC^{N \times N})}.
\]
A {\em band operator} on $l^p(\sZ^n, \sC^N)$ is an operator of the 
form
\begin{equation} \label{0.1}
A = \sum_{|\alpha| \le m} a_\alpha V_\alpha
\end{equation}
with coefficients $a_\alpha \in l^\infty (\sZ^n, \, \sC^{N \times N})$. 
The closure in $\cL (l^p(\sZ^n, \, \sC^N))$ of the set of all band 
operators is a subalgebra of $\cL (l^p(\sZ^n, \sC^N))$. We denote this 
algebra by $\cA(l^p(\sZ^n, \, \sC^N))$ and call its elements 
{\em band-dominated operators} (BDO for short). In a completely 
analogous way, band-dominated operators on $l^\infty (\sZ^n, \, \sC^N)$ 
are defined.

Our main tool to study Fredholm properties of band-dominated
operators are the associated limit operators.  
\begin{definition} \label{def0.1}
Let $A \in \cL (l^p(\sZ^n, \, \sC^N))$, and let $h : \sN \to \sZ^n$ 
be a sequence tending to infinity. A linear operator $A_h$ is called 
the limit operator of $A$ with respect to the sequence $h$ if
\[
V_{-h(m)} A V_{h(m)} \to A_h \quad \mbox{and} \quad V_{-h(m)} A^*
V_{h(m)} \to A_h^*
\]
strongly as $m \to \infty$. We let $\opp A$ denote the set of all
limit operators of $A$.
\end{definition}
Here and in what follows, convergence of a sequence in $\sZ^n$ to 
infinity means convergence of this sequence to infinity in the 
one-point compactification of $\sZ^n$ (which makes sense since 
$\sZ^n$ is a locally compact metric space).

There are operators on $l^p(\sZ^n, \, \sC^N)$ which do not possess 
limit operators at all. But if $A$ is a band-dominated operator then 
one can show via a Cantor diagonal argument that {\em every} sequence
$h$ tending to infinity has a subsequence $g$ for which the limit
operator $A_g$ exists. Moreover, the operator spectrum of $A$ stores
the complete information on the Fredholmness of $A$, as the
following theorem states. (In case $n=1$ there is also a sufficiently
nice formula for the Fredholm index of $A$ which expresses this index 
in terms of local indices of the limit operators of $A$, see 
\cite{RRR1}.)
\begin{theorem} \label{te0.1}
An operator $A \in \cA(l^p(\sZ^n, \, \sC^N))$ is Fredholm if and only 
if all limit operators of $A$ are invertible and if
\begin{equation} \label{0.1'}
\sup_{A_h \in \opp (A)} \|A_h^{-1}\| < \infty.
\end{equation}
\end{theorem}
The uniform boundedness condition (\ref{0.1'}) is often difficult to
check: It is one thing to verify the invertibility of an operator
and another one to provide a good estimate for the norm of its
inverse. It is therefore of vital importance to single out classes
of band-dominated operators for which this condition is
automatically satisfied. One of these classes is defined by imposing
conditions of the decay of the norms of the coefficients. More
precisely, we consider band-dominated operators of the form
\[
A := \sum_{\alpha \in \sZ^n} a_\alpha V_\alpha
\]
where
\begin{equation} \label{0.2}
\sum_{\alpha \in \sZ^n} \|a_\alpha \|_{l^\infty(\sZ^n, \, 
\sC^{N \times N})} < \infty.
\end{equation}
One can show that the set $W(\sZ^n, \sC^N)$ of all operators with 
property (\ref{0.2}) forms an algebra and that the term on the 
left-hand side of (\ref{0.2}) defines a norm which makes 
$W(\sZ^n, \, \sC^N)$ to a Banach algebra. We refer to this algebra 
as the {\em Wiener algebra} and write $\|A\|_{W(\sZ^n, \, \sC^N)}$ 
for the norm of an operator in $W(\sZ^n, \, \sC^N)$. Clearly, 
operators in the Wiener algebra act boundedly on each of the 
spaces $l^p(\sZ^n, \, \sC^N)$ (including $p = \infty$) and
\[
\|A\|_{\cL(l^p(\sZ^n, \, \sC^N))} \le \|A\|_{W(\sZ^n, \, \sC^N)}.
\]
Hence, $W(\sZ^n, \, \sC^N) \subseteq \cA(l^p(\sZ^n, \, \sC^N))$ for 
every $p$.

One important property of the Wiener algebra is its inverse
closedness in each of the algebras $\cL(l^p(\sZ^n, \, \sC^N))$, 
i.e., if $A \in W(\sZ^n, \, \sC^N)$ has an inverse in $\cL(l^p 
(\sZ^n, \, \sC^N))$, then $A^{-1}$ belongs to $W(\sZ^n, \, \sC^N)$ 
again. This fact implies that the spectrum of an operator $A \in 
W(\sZ^n, \, \sC^N)$ considered as acting on $l^p(\sZ^n, \, \sC^N)$
does not depend on $p \in (1, \, \infty)$. Also the operator 
spectrum $\opp (A)$ proves to be independent of $p$, which justifies 
to write $\op A$ instead. Note finally that all limit operators of 
operators in the Wiener algebra belong to the Wiener algebra again.

For operators in the Wiener algebra, the Fredholm criterion in
Theorem \ref{te0.1} reduces to the following much simpler
assertion.
\begin{theorem} \label{te0.2}
Let $A \in W(\sZ^n, \, \sC^N)$. The operator $A$ is Fredholm on 
$l^p(\sZ^n, \, \sC^N)$ if and only if there exists a $p_0 \in 
[1, \, \infty]$ such that all limit operators of $A$ are invertible 
on $l^{p_0} (\sZ^n, \, \sC^N)$.
\end{theorem}
Theorem \ref{te0.2} has the following useful consequence.
\begin{theorem} \label{te0.3}
For $A \in W(\sZ^n, \, \sC^N)$, the essential spectra of
$A : l^p(\sZ^n, \, \sC^N) \to l^p(\sZ^n, \, \sC^N)$ do not 
depend on $p \in (1, \, \infty)$, and 
\begin{equation} \label{0.3}
\spe A = \bigcup_{A_h \in \op A} \spec A_h.
\end{equation}
\end{theorem}
\section{BDO on periodic discrete metric spaces} \label{sect3}
\subsection{Periodic discrete metric spaces}
By a {\em discrete metric space} we mean a countable set $X$
together with a metric $\rho$ such that every ball
\[
B_r (x_0) := \{ x \in X : \rho (x, \, x_0) \le r \}
\]
is a finite set. For each discrete metric space $X$, we introduce
some standard Banach spaces over $X$. For $p \in (1, \, \infty)$, 
let $l^p(X)$ denote the Banach space of all complex-valued functions 
$u$ on $X$ with norm
\[
\|u\|_{l^p (X)}^p := \sum_{x \in X} |u(x)|^p,
\]
and write $l^\infty (X)$ for the Banach space of all bounded functions 
$u$ of $X$ with norm 
\[
\|u\|_{l^\infty (X)} :=  \sup_{x \in X} |u(x)|
\]
A {\em periodic discrete metric space} is a discrete metric space
provided with the free action of the group $\sZ^n$. More
precisely, let $X$ be a discrete metric space, and let there be a
mapping
\[
\sZ^n \times X \to X, \quad (\alpha, \, x) \to \alpha \cdot x
\]
satisfying
\[
0 \cdot x = x \quad \mbox{and} \quad (\alpha + \beta) \cdot x =
\alpha \cdot (\beta \cdot x)
\]
for arbitrary elements $\alpha, \, \beta \in \sZ^n$ and $x \in X$,
which leaves the metric invariant,
\begin{equation} \label{1.1'}
\rho (\alpha \cdot x, \, \alpha \cdot y) = \rho (x, \, y)
\end{equation}
for all elements $\alpha \in \sZ^n$ and $x, \, y \in X$. Recall 
also that the group $\sZ^n$ acts {\em freely} on $X$ if whenever 
the equality $x = \alpha \cdot x$ holds for elements $x \in X$ 
and $\alpha \in \sZ^n$ then, necessarily, $\alpha = 0$. 

For each element $x \in X$, consider its orbit $\{\alpha \cdot x
\in X : \alpha \in \sZ^n \}$ with respect to the action of
$\sZ^n$. Any two orbits are either disjoint or identical.
Hence, there is a binary equivalence relation on $X$, by calling
two points equivalent if they belong to the same orbit. The set of
all orbits of $X$ with respect to the action of $\sZ^n$ is denoted
by $X/\sZ^n$. A basic assumption throughout what follows is that
the orbit space $X/\sZ^n$ is {\em finite}. Thus, there is a finite
subset $\cM := \{x_1, \, x_2, \, \ldots, \, x_N\}$ of $X$ such
that the orbits
\[
X_j := \{ \alpha \cdot x_j \in X : \alpha \in \sZ^n \}
\]
satisfy $X_i \cap X_j = \emptyset$ if $x_i \neq x_j$ and
$\cup_{i=1}^N X_i = X$. If all these conditions are satisfied 
then we call $X$ is a {\em periodic discrete metric space} 
with respect to $\sZ^n$ or simply $\sZ^n$-periodic.   

The free action of $\sZ^n$ on $X$ guarantees that the mapping
\[
U_j : \sZ^n \to X_j, \quad \alpha \mapsto \alpha \cdot x_j
\]
is a bijection for every $j = 1, \, \ldots, \, N$. For each 
complex-valued function $f$ on $X$, let $Uf : \sZ^n \to \sC^N$ 
be the function
\[
(Uf)(\alpha) := ((U_1 f)(\alpha), \, \ldots, \, (U_N f) (\alpha)).
\]
Clearly, the mapping $U$ is a linear isometry from $l^p(X)$ onto
$l^p(\sZ^n, \, \sC^N)$, and the mapping $A \mapsto U A U^{-1}$ is 
an isometric isomorphism from $\cL(l^p(X))$ onto $\cL(l^p(\sZ^n, \,
\sC^N))$ for every $p \in [1, \, \infty]$.

Another consequence of our assumptions is that 
\begin{equation} \label{1.1''}
\lim_{\sZ^n \ni \alpha \to \infty} \rho (\alpha \cdot x, \, y) =
\infty.
\end{equation}
for all points $x, \, y \in X$. Indeed, suppose that (\ref{1.1''}) 
is wrong. Then there are points $x, \, y \in X$, a positive constant 
$M$, and a sequence $\alpha$ of pairwise different points in $\sZ^n$ 
such that  
\begin{equation} \label{erg}
\rho (\alpha(n) \cdot x, \, y) \le M \quad \mbox{for all} \; n \in 
\sN.
\end{equation}
The free action of $\sZ^n$ on $X$ implies that $(\alpha(n) \cdot 
x)_{n \in \sN}$ is a sequence of pairwise different points in $X$.
Hence, (\ref{erg}) implies that the ball with center $y$ and radius 
$M$ contains infinitely many points, a contradiction. \hfill \qed
\subsection{Band-dominated operators on $X$}
Let $X$ be a periodic discrete metric space and $p \in [1, \,
\infty)$. We consider linear operators $A$ on $l^p(X)$ for which
there exists a function $k_A \in l^\infty (X \times X)$ such that
\begin{equation} \label{1.1}
(Au)(x) = \sum_{y \in X} k_A (x, \, y) u(y) \quad \mbox{for all}
\; x \in X
\end{equation}
and for all finitely supported functions $u$ on $X$ (note that the
latter form a dense subspace of $l^p(X)$). The function $k_A$ is
called the {\em generating function} of the operator $A$. It is
easily seen that every bounded operator $A$ on $l^p(X)$ is of this
form and is, thus, generated by a bounded function. The converse
is certainly not true. It is also clear that every operator $A$ 
determines its generating function uniquely, since
\[
(A \delta_y)(x) = k_A (x, \, y)
\]
where $\delta_y$ is the function on $X$ which is 1 at $y$ and 0 at
all other points.

An operator $A$ of the form (\ref{1.1}) is called a {\em band
operator} if there exists an $R > 0$ such that $k_A(x, \, y) = 0$
whenever $\rho (x, \, y) > R$.
\begin{example}
Every operator $aI$ of multiplication by a function $a \in
l^\infty (X)$ is a band operator. \hfill \qed
\end{example}
\begin{example}
For $\alpha \in \sZ^n$, let $T_\alpha$ be the operator of shift by
$\alpha$ on $l^p (X)$, i.e., $(T_\alpha u)(x) := u((- \alpha)
\cdot x)$. Clearly, $T_\alpha$ is a band operator which acts as an
isometry on $l^p (X)$. Hence, every operator of the form
\begin{equation} \label{e1.2}
\sum_{|\alpha| \le m} a_\alpha T_\alpha
\end{equation}%
with $a_\alpha \in l^\infty (X)$ is a band operator (but there are
band operators which can not be represented of this form). \hfill 
\qed
\end{example}
\begin{proposition} \label{p3.2'}
If $A$ is a band operator on $l^p(X)$, then $UAU^{-1}$ is a band
operator on $l^p(\sZ^n, \, \sC^N)$.
\end{proposition}
\begin{proof}
The operator $UAU^{-1}$ has the matrix representation
\begin{equation} \label{e1.1}
(UAU^{-1} f)_i (\alpha) = \sum_{j=1}^N \sum_{\beta \in \sZ^n}
r_A^{ij} (\alpha, \, \beta) f_j(\beta)
\end{equation}
where $\alpha \in \sZ^n$, $i = 1, \, \ldots, \, N$ and
\begin{equation} \label{e3.1}
r_A^{ij} (\alpha, \, \beta) := k_A (\alpha \cdot x_i, \, \beta
\cdot x_j).
\end{equation}
From (\ref{1.1''}) we conclude that 
\[
\rho(\alpha \cdot x_i, \, \beta \cdot x_j) = 
\rho (x_i, \, (\beta - \alpha) \cdot x_j) \to \infty
\]
as $|\alpha - \beta| \to \infty$. Thus, there is an $R_1 > 0$ such 
that $r_A^{ij} (\alpha, \, \beta) = 0$ if $|\alpha - \beta| > R_1$.
In other words, every $r_A^{ij}$ is the generating function of a 
band operator on $l^p(\sZ^n)$, implying that $UAU^{-1}$ is a band
operator on $l^p(\sZ^n, \, \sC^N)$.
\end{proof} \\[3mm]
The preceding proposition implies in particular that every band
operator is bounded on $l^p (X)$ for $p \in [1, \, \infty]$.

For $p \in [1, \, \infty]$, let $\cA_p (X)$ stand for the closure
in $\cL(l^p(X))$ of the set of all band operators. The operators
in $\cA_p (X)$ are called {\em band-dominated operators on} $X$.
Note that the class $\cA_p (X)$ depends heavily on $p$ (whereas
the class of the band operators is independent of $p$). One can
show easily (for example, by employing the preceding proposition
and the well properties of band-dominated operators on $\sZ^n$)
that $\cA_p (X)$ is a Banach algebra and even a $C^*$-algebra if
$p=2$. 
\begin{proposition} \label{pr1.1}
Let $X$ be a periodic discrete metric space and $p \in [1, \,
\infty]$. The mapping $A \mapsto U A U^{-1}$ is an isomorphism
between the Banach algebras $\cA_p (X)$ and $\cA_p (\sZ^n, \,
\sC^N)$.
\end{proposition}
\begin{proof}
Note that an operator $A$ is a band operator on $l^p (X)$ if and
only if $U A U^{-1}$ is a band operator on $l^p(\sZ^n, \, \sC^N)$. The
assertion follows since the mapping $A \mapsto U A U^{-1}$ is a
continuous isomorphism between the Banach algebras $\cL(l^p (X))$
and $\cL(l^p(\sZ^n, \, \sC^N))$. 
\end{proof}
\subsection{Limit operators and Fredholmness}
Let $X$ be a $\sZ^n$-periodic discrete metric space. The goal of 
this section is a criterion for the Fredholmness of band-dominated 
operators on $l^p(X)$. This criterion makes use of the limit 
operators of $A$ which, in a sense, reflect the behaviour of $A$ 
at infinity. Here is the definition.
\begin{definition} \label{d1.1 copy(1)}
Let $1 < p < \infty$, and $h : \sN \to \sZ^n$ be a sequence
tending to infinity. We say that $A_h$ is a {\em limit operator}
of $A \in \cL(l^p(X))$ defined by the sequence $h$ if
\[
T_{h(m)}^{-1} A T_{h(m)} \to A_h \quad \mbox{and} \quad
T_{h(m)}^{-1} A^* T_{h(m)} \to A_h^* \quad \mbox{as} \; m \to
\infty
\]
strongly on $l^p(X)$ and $l^p(X)^* = l^q(X)$ with $1/p + 1/q = 1$,
respectively. We denote the set of all limit operators of $A \in
\cL(l^p(X))$ by $\opp (A)$ and call this set the {\em operator
spectrum} of $A$.
\end{definition}
Note that the generating function of the shifted operator
$T_\alpha^{-1} A T_\alpha$ is related with that of $A$ by
\begin{equation} \label{equa0}
k_{T_\alpha^{-1} A T_\alpha} (x, \, y) = k_A ((- \alpha) \cdot x,
\, (- \alpha) \cdot y)
\end{equation}
and that the generating functions of $T_{h(m)}^{-1} A T_{h(m)}$
converge point-wise on $X \times X$ to the generating function of
the limit operator $A_h$ if the latter exists.

It is an important property of band-dominated operators that their
operator spectrum is not empty. More general, one has the following
result which can be proved by an obvious Cantor diagonal argument 
(see \cite{RRS1,RRS2,RRSB}).
\begin{proposition} \label{prop1}
Let $p \in (1, \, \infty)$ and $A \in \cA_p (X)$. Then
every sequence $h : \sN \to G$ which tends to infinity possesses a
subsequence $g$ such that the limit operator $A_g$ of $A$ with
respect to $g$ exists.
\end{proposition}
The following theorem settles the basic relation between the
Fredholmness of a band-dominated operator $A$ and the
invertibility of its limit operators. It follows easily from
Theorem \ref{te0.1} if one takes into account that the mapping
\[
\cA_p (X) \to \cA_p (\sZ^n, \, \sC^N), \quad A \mapsto UAU^{-1}
\]
is an isomorphism of Banach algebras and that the relation
\[
(UAU^{-1})_h = U A_hU^{-1}
\]
between the limit operators of $A$ and $UAU^{-1}$ holds.
\begin{theorem} \label{t1.1}
Let $p \in (1, \, \infty)$ and $A \in \cA_p(X)$. Then $A$ is a
Fredholm operator on $l^p(X)$ if and only if all limit operators
of $A$ are invertible and if the norms of their inverses are
uniformly bounded,
\begin{equation} \label{1.3}
\sup_{A_h \in op(A)} \|A_h^{-1}\| < \infty.
\end{equation}
\end{theorem}
\subsection{The Wiener algebra of $X$}
The goal of this section is to single out a class of
band-dominated operators for which the uniform boundedness
condition (\ref{1.3}) is redundant.
\begin{definition}
Let $X$ be a $\sZ^n$-periodic discrete metric space. The set 
$\cW(X)$ consists of all linear operators $A$ for which there 
is a function $h_A$ in $l^1(\sZ^n)$ such that
\begin{equation} \label{1.2}
\max_{j \in \{1, \, \ldots, \, N\}} \sum_{i=1}^N |r_A^{ij}(\alpha,
\, \beta)| \le h_A (\alpha - \beta)
\end{equation}
for all $\alpha, \, \beta \in \sZ^n$.
\end{definition}
We introduce a norm in $\cW(X)$ by
\begin{equation} \label{1.2'}
\|A\|_{\cW(X)} := \inf \|h\|_{l^1(\sZ^n)}
\end{equation}
where the infimum is taken over all sequences $h \in l^1(\sZ^n)$
for which inequality (\ref{1.2}) holds in place of $h_A$.
\begin{proposition} \label{p3.2}
The set $\cW(X)$ with the norm (\ref{1.2'}) is a Banach algebra,
and the mapping $A \mapsto UAU^{-1}$ is an isometrical isomorphism
between the Banach algebras $\cW(X)$ and $\cW(\sZ^n, \, \sC^N)$.
\end{proposition}
The proof is straightforward. We refer to the algebra $\cW(X)$ as
the {\em Wiener algebra}.
\begin{proposition} \label{p3.3}
Let $p \in [1, \, \infty]$. \\[1mm]
(i) Every operator $A \in \cW(X)$ is bounded on each of the spaces
$l^p(X)$. \\[1mm]
(ii) The algebra $\cW(X)$ is inverse closed in each of the
algebras $\cL(l^p(X))$.
\end{proposition}
Proposition \ref{p3.3} follows from Proposition \ref{p3.2} and the
related results for the special case $X = \sZ^n$ presented in
\cite{RRS1,RRS2} and \cite{RRSB}. \hfill \qed \\[3mm]
The following result highlights the importance of the Wiener
algebra in our context.
\begin{theorem} \label{t1.2}
Let $A \in \cW(X)$. Then $A$ is a Fredholm operator on $l^p(X)$
with $p \in (1, \, \infty)$ if and only if there is a $p_0 \in [1,
\, \infty]$ such that all limit operators of $A$ are invertible on
$l^{p_0} (X)$. Moreover $\spe A$ does not depend on $p \in (1, \,
\infty)$, and
\begin{equation} \label{1.5''}
\spe A = \bigcup_{A_h \in \op (A)} \spec A_h.
\end{equation}
\end{theorem}
Theorem \ref{t1.2} follows immediately from Proposition \ref{p3.2}
and Theorems \ref{te0.2} and \ref{te0.3}.

The following result states a sufficient condition for the
absence of the discrete spectrum of an operator $A \in \cA_p (X)$.
\begin{proposition} \label{p1.4}
Let $A \in \cA_p(X)$ and suppose there is a sequence $h : \sN \to
\sZ^n$ for which the limit operator $A_h$ exists in the sense of
norm convergence,
\begin{equation} \label{1.6'}
\lim_{m \to \infty} \|T_{h_m}^{-1} A T_{h_m} - A_h \| = 0.
\end{equation}
Then $\spe A = \spec A$.
\end{proposition}
\begin{proof}
Let $\lambda \notin \spe A$. Then, by Theorem \ref{t1.1}, $\lambda
\notin \spec A_h$. It follows from (\ref{1.6'}) that $\lambda
\notin \spec A$. Hence, $\spec A \subseteq \spe A$, which implies
the assertion.
\end{proof}
\section{Periodic operators on periodic metric spaces} \label{sect4}
Let $X$ be a $\sZ^n$-periodic discrete metric space. An operator 
$A \in \cL(l^p(X))$ is said to be {\em $\sZ^n$-periodic} if it is 
invariant with respect to left shifts by elements of $\sZ^n$, that 
is if
\[
T_\alpha A = A T_\alpha \qquad \mbox{for every} \; \alpha \in
\sZ^n.
\]
The following is a straightforward consequence of Proposition
\ref{p1.4}.
\begin{proposition} \label{pr2.1}
Let $A \in \cA_p (X)$ be a $\sZ^n$-periodic operator. Then
\[
\spe A = \spec A.
\]
\end{proposition}
The explicit description of the spectrum (= the essential
spectrum) of $\sZ^n$-periodic operators is possible by means 
of the Fourier transform. One easily checks that $A \in \cW(X)$ is 
$\sZ^n$-periodic on $X$ if and only if the generating function
$k_A$ of $A$ satisfies the following periodicity condition: For
all group elements $\gamma \in \sZ^n$ and all points $x, \, y \in
X$,
\[
k_A (\gamma \cdot x, \, \gamma \cdot y) = k_A(x, \, y).
\]
This equality implies that the functions $r_A^{ij}(\alpha, \,
\beta) := k_A (\alpha \cdot x_i, \, \beta \cdot x_j)$ satisfy
\[
r_{A}^{ij} (\alpha, \, \beta) = k_A ((\alpha - \gamma) \cdot x_i,
\, (\beta - \gamma) \cdot x_j)
\]
for all $\gamma \in \sZ^n$, whence $r_A^{ij}(\alpha, \, \beta) =
r_A^{ij} (\alpha - \beta, \, 0)$. Hence, for $i = 1, \, \ldots, \,
N$,
\begin{eqnarray*}
(U A U^{-1} f)_i (\alpha ) & = & \sum_{j=1}^N \sum_{\beta \in \sZ^n}
r_A^{ij}(\alpha, \, \beta) \, (U_j f)(\beta) \\
& = & \sum_{j=1}^N \sum_{\beta \in \sZ^n} r_A^{ij}(\alpha - \beta,
\, 0) \, (U_j f)(\beta) \\
& = & \sum_{j=1}^N \sum_{\beta \in \sZ^n} r_A^{ij}(\beta, \, 0) \,
(V_\beta U_j f)(\alpha)
\end{eqnarray*}
where
\[
|r_A^{ij} (\beta, \, 0)| \le h (\beta)
\]
for a some non-negative function $h \in l^1(\sZ^n)$. Thus, we
arrived at the following proposition.
\begin{proposition} \label{pr2.2}
Every $\sZ^n$-periodic operator $A \in \cW(X)$ is isometrically 
equivalent to the shift invariant matrix operator $U A U^{-1} 
\in W(\sZ^n, \, \sC^N)$.
\end{proposition}
Under the conditions of the previous proposition, we associate
with $A$ a function $\sigma_A : \sT^n \to \sC^{N \times N}$ via
\[
\sigma_A (t) := \sum_{\beta \in \sZ^n} r_A (\beta) \, t^\beta
\]
where $\sT$ is the torus $\{z \in \sC : |z|=1 \}$, $r_A (\beta)$
is the matrix $(r_A^{ij}(\beta, \, 0))_{i, \, j=1}^N$, and
$t^\beta := t_1^{\beta_1} \ldots t_n^{\beta_n} $ for $t = (t_1, \,
\ldots, \, t_n) \in \sT^n$ and $\beta = (\beta_1, \, \ldots, \,
\beta_n) \in \sZ^n$. The function $\sigma_A$ is referred to as the
{\em symbol} of $A$. It is well known that the operator
\[
(\tilde{A}u) (\alpha) := \sum_{\beta \in \sZ^n} r_A(\alpha -\beta,
\,  0) u(\beta)
\]
is invertible on $l^p(\sZ^n, \, \sC^N)$ with $p \in [1, \,
\infty]$ if and only if $\det \sigma_A \neq 0$ on $\sT^n$.

For $t \in \sT^n$, let $\lambda_A^j (t)$ with $j = 1, \, \ldots,
\, N$ denote the eigenvalues of the matrix $\sigma_{A}(t)$. The
enumeration of the eigenvalues can be chosen in such a way that
$\lambda_A^j (t)$ depends continuously on $t$ for every $j$. Thus,
the sets
\begin{equation} \label{1.5'}
\cC_j (A) := \{ \lambda \in \sC : \lambda = \lambda_A^j(t), \, t
\in \sT^n \}, \quad j=1, \, \ldots, \, N
\end{equation}
are compact and connected curves in the complex plane, called the
{\em spectral} or {\em dispersion curves} of $A$.
\begin{proposition} \label{p1.3}
Let $A \in \cW(X)$ be a $\sZ^n$-periodic operator. Then
\begin{equation} \label{1.6}
\spec A = \spe A = \bigcup_{j=1}^N \cC_j (A).
\end{equation}
\end{proposition}
If, moreover, $A \in \cW(X)$ is a self-adjoint 
$\sZ^n$-periodic operator on $l^2 (X)$, then $\sigma_A$ is a 
Hermitian matrix-valued function. Hence, the $\lambda_A^j$ are
continuous real-valued functions, and
\[
\cC_j(A) = [\alpha_j(A), \, \beta_j(A)] \quad \mbox{for} \; j =
1, \, \ldots , \, N
\]
where $\alpha_j (A) := \min_{t\in \sT^n} \lambda_A^j(t)$ and
$\beta_j(A) := \max_{t \in \sT^n} \lambda_A^j(t)$. Thus, the
spectrum of a self-adjoint $\sZ^n$-periodic operator on a 
periodic metric space is the union of at most $N$ compact 
intervals (with $N$ the number of orbits of $X$ under the 
action of $\sZ^n$).
\section{Operators with slowly oscillating coefficients on periodic
metric spaces} \label{sect5}
Let again $X$ be a $\sZ^n$-periodic discrete metric space. A 
function $a \in l^\infty (X)$ is called {\em slowly oscillating} 
if, for every two points $x, \, y \in X$,
\begin{equation} \label{2.1}
\lim_{\alpha \to \infty} (a(\alpha \cdot x) - a(\alpha \cdot y)) =
0.
\end{equation}
The set of all slowly oscillating functions on $X$ forms a
$C^*$-subalgebra of $l^\infty (X)$ which we denote by $SO(X)$.
Note that the class $SO(X)$ does not only depend on $X$ but 
also on the action of $\sZ^n$ on $X$.

Let $a \in SO(X)$ and $h : \sN \to G$ be a sequence tending to
infinity. The Bolzano-Weierstrass Theorem and a Cantor diagonal
argument imply that there is a subsequence $g$ of $h$ such that
the functions $x \mapsto a(g(m) \cdot x)$ converge point-wise to a
function $a_g \in l^\infty (X)$ as $m \to \infty$. The condition
(\ref{2.1}) ensures that the limit function $a_g$ is $\sZ^n$-periodic 
on $X$. Indeed, for every $\alpha \in \sZ^n$,
\[
a_g (x) - a_g (\alpha \cdot x) = \lim_{m \to \infty} (a(g(m) \cdot
x) - (a(g(m) \cdot (\alpha \cdot x))) = 0.
\]
We consider the operators of the form
\begin{equation} \label{2.2}
A = \sum_{k, \, l=1}^\infty b_k \, A_{kl} \, c_lI
\end{equation}
where the $A_{kl}$ are $\sZ^n$-periodic operators in $\cW(X)$ and the
$b_k$ and $c_l$ are slowly oscillating functions satisfying
\[
\sum_{k, \, l=1}^\infty \|b_k\|_{l^\infty (X)} \,
\|A_{kl}\|_{\cW(X)} \, \|c_l\|_{l^\infty (X)} < \infty.
\]
Let $h : \sN \to \sZ^n$ be a sequence tending to infinity. Then
\[
T_{h(m)}^{-1} A T_{h(m)} = \sum_{k, \, l=1}^\infty (T_{h(m)}^{-1}
b_k) \, A_{kl} \, (T_{h(m)}^{-1} c_l)I.
\]
One can assume without loss that the point-wise limits
\[
\lim_{m \to \infty} (T_{h(m)}^{-1} b_k)(x) =: b_k^h, \qquad
\lim_{m \to \infty} (T_{h(m)}^{-1} c_l)(x) =: c_l^h
\]
exist (otherwise we pass to a suitable subsequence of $h$). As we
have seen above, the limit functions $b_k^h$ and $c_l^h$ are
$\sZ^n$-periodic on $X$. Consequently, the limit operators $A_h$ 
of $A$ are $\sZ^n$-periodic operators of the form
\[
A_h = \sum_{k, \, l=1}^\infty b_k^h \, A_{kl} \, c_l^hI.
\]
Now, the following is an immediate consequence of Theorem
\ref{t1.2}.
\begin{theorem} \label{t2.1}
Let $A$ be an operator with slowly oscillating coefficients of the
form $(\ref{2.2})$. Then $A$ is a Fredholm operator on $l^p(X)$ if
and only if, for every operator $A_h \in \op A$,
\[
\det \sigma_{A_h} (t) \neq 0 \quad \mbox{for every} \; t \in \sT^n.
\]
Moreover,
\[
\spe A = \bigcup_{A_h \in \op (A)} \spec A_h = \bigcup_{A_h \in
\op (A)} \bigcup_{j=1}^N \cC_j (A_h).
\]
\end{theorem}
\section{Schr\"{o}dinger operators on periodic graphs} \label{sect6}
By a {\em discrete infinite graph} we mean a countable set $X$
together with a binary relation $\sim$ which is anti-reflexive
(i.e., there is no $x \in X$ such that $x \sim x$) and symmetric
and which has the property that for each $x \in X$ there are only
finitely many $y \in X$ such that $x \sim y$. The points of $X$
are called the {\em vertices} and the pairs $(x, \, y)$ with $x
\sim y$ the {\em edges} of the graph. Due to anti-reflexivity,
the graphs under consideration do not possess loops. We write
$m(x)$ for the number of edges starting (or ending) at the vertex
$x$ of $X$. If $x \sim y$, we say that the vertices $x, \, y$ are
{\em adjacent}.

For technical reasons it will be convenient to assume that the
graph $(X, \, \sim)$ is connected, i.e., given distinct points $x,
\, y \in X$, there are finitely many points $x_0, \, x_1, \,
\ldots, x_n \in X$ such that $x_0 = x$, $x_n = y$ and $x_i \sim
x_{i+1}$ for $i = 0, \, \ldots, \, n$. The smallest number $n$
with this property defines the {\em graph distance} $\rho(x, \,
y)$ of $x$ and $y$. Together with $\rho(x, \, x) := 0$, this
defines a metric $\rho$ on $X$ which makes $X$ to discrete 
metric space.

We call $(X, \, \sim)$ a {\em $\sZ^n$-periodic discrete graph} 
if it is a connected discrete infinite graph, if the group $\sZ^n$ 
operates freely from the left on $X$, and if the group action 
respects the graph structure, i.e.,
\[
x \sim y \qquad \mbox{if and only if} \qquad \alpha \cdot x \sim
\alpha \cdot y
\]
for arbitrary vertices $x, \, y \in X$ and group elements $\alpha
\in \sZ^n$. Clearly, every group with these properties leaves the
graph distance invariant, that is, $X$ becomes a $\sZ^n$-periodic 
discrete metric space. If $(X, \, \sim)$ is a $\sZ^n$-periodic 
graph, then the function $m$ is $\sZ^n$-periodic, too, that is,
$m(\alpha \cdot x) = m(x)$ for every $x \in X$ and $\alpha \in
\sZ^n$.

Every $\sZ^n$-periodic discrete graph $\Gamma := (X, \, \sim)$ 
induces a canonical difference operator $\Delta_\Gamma$ on $l^p(X)$, 
called the (discrete) {\em Laplace operator} or {\em Laplacian} of 
$\Gamma$, via 
\begin{equation} \label{3.1}
(\Delta_\Gamma u)(x) := \frac{1}{m(x)} \sum_{y \sim x} u(y), \quad
x \in X.
\end{equation}
Evidently, $\Delta_\Gamma$ is a $\sZ^n$-periodic band operator.

Let $v \in l^\infty (X)$. The operator $\cH_\Gamma :=
\Delta_\Gamma + vI$ is referred to as the (discrete) {\em
Schr\"{o}dinger operator with electric potential} $v$ on the graph
$X$. Given a sequence $h : \sN \to \sZ^n$ tending to infinity,
there exist a subsequence $g$ of $h$ and a function $v^g \in
l^\infty (X)$ such that $v(g(m) \cdot x) \to v^g (x)$ as $m \to
\infty$ for every $x \in X$. It turns out that the operator
\[
\cH_\Gamma^g := \Delta_\Gamma + v^g I
\]
is the limit operator of $\cH_\Gamma$ defined by the sequence $g$
and that every limit operator of $\cH_\Gamma$ is of this form.
Thus, Theorem \ref{t1.2} implies the following.
\begin{theorem} \label{t3.1}
The Schr\"{o}dinger operator $\cH_\Gamma = \Delta_\Gamma + vI$
with bounded potential $v$ is a Fredholm operator on $l^p(X)$ with
$p \in (1, \, \infty)$ if and only if there is a $p_0 \in 
[1, \, \infty]$ such that all limit operators of $\cH_\Gamma$ 
are invertible on $l^{p_0}(X)$. The essential spectrum of 
$\cH_\Gamma$ does not depend on $p \in (1, \, \infty)$, and
\begin{equation} \label{3.2}
\spe \cH_\Gamma = \bigcup_{\cH_\Gamma^h \in \op (\cH_\Gamma)}
\spec \cH_\Gamma^h.
\end{equation}
\end{theorem}
For an explicit description of the essential spectrum of the
Schr\"{o}dinger operator $\cH_\Gamma$ we first assume that $v$ is
a periodic potential. Then the operator $UvU^{-1}$ is the operator
of multiplication by the diagonal matrix $\diag (v(x_1), \,
\ldots, \, v(x_N))$. Hence,
\[
U \cH_\Gamma U^{-1} = \sum_{\alpha \in \{-1, \, 0, \, 1\}^n}
a_\alpha V_\alpha + \diag (v(x_1), \, \ldots, \, v(x_N)),
\]
where the $a_\alpha$ are certain constant $N \times N$ matrices
which depend on the structure of the graph $\Gamma$. Consequently,
\[
\sigma_{\cH_\Gamma}(t) = \sum_{\alpha \in \{-1, \, 0, \, 1\}^n}
a_\alpha t^\alpha + \diag (v(x_1), \, \ldots, \, v(x_N)), \quad t
\in \sT^n.
\]
If the potential $v$ is real-valued, then $\cH_\Gamma$ acts as a
self-adjoint operator on $l^2(X)$, and $\sigma_{\cH_\Gamma}$ is a
Hermitian matrix-valued function on $\sT^n$. From Proposition
\ref{p1.3} we conclude that
\[
\spec \cH_\Gamma = \bigcup_{j=1}^N \cC_j (\cH_\Gamma)
\]
where $\cC_j (\cH_\Gamma)$ is the real interval $[a_j, \, b_j]$
with $a_j := \min_{t \in \sT^n} \lambda_{\cH_\Gamma}^j(t)$ and
$b_j := \max_{t \in \sT^n} \lambda_{\cH_\Gamma}^j(t)$.

Next we consider Schr\"{o}dinger operators $\cH_\Gamma =
\Delta_\Gamma + vI$ with slowly oscillating potential $v$. As we
have seen in the previous section, all limit operators of
$\cH_\Gamma$ are of the form
\[
\cH_\Gamma^g = \Delta_\Gamma + v^g I
\]
with periodic potentials $v^g$. Theorem \ref{t3.1} together with
Theorem \ref{t1.2} yield the following.
\begin{theorem}
Let $\cH_\Gamma = \Delta_\Gamma + vI$ with $v \in SO(X)$. Then
\[
\spe \cH_\Gamma = \bigcup_{\cH_\Gamma^g \in \op (\cH_\Gamma)}
\bigcup_{j=1}^N \cC_j (\cH_\Gamma^g)
\]
with the spectral curves $\cC_j (\cH_\Gamma^g)$ defined as in
$(\ref{1.5'})$.
\end{theorem}
If the slowly oscillating potential $v$ is real-valued, then the
spectral curves $\cC_j (\cH_\Gamma^g)$ are (possibly overlapping) 
intervals on the real line.

The following examples clarify the structure of the essential
spectrum of Schr\"{o}dinger operators on some special periodic
graphs. The graphs under consideration are embedded into 
$\sR^n$ for some $n$. This embedding allows one to consider the 
vertices of the graph as vectors and to use the linear structure 
of $\sR^n$ in order to describe the group action.
\begin{example}[The Cayley graph of $\sZ^n$] \label{ex-1}
As every finitely generated group, the group $\sZ^n$ induces a
graph (called the Cayley graph of the group) the vertices of which
are the points in $\sZ^n$ and with edges $(\alpha, \, \alpha \pm
e_i)$ where $\alpha \in \sZ^n$ and where $e_i := (0, \, \ldots, \,
0, \, 1, 0, \, \ldots, \, 0)$ with the 1 at the $i$th position and
$i = 1, \, \ldots, \, n$. The Laplace operator $\Delta_{\sZ^n}$ is
of the form
\[
(\Delta_{\sZ^n} u)(x) = \frac{1}{2n} \sum_{i=1}^n (u(x + e_i) +
u(x - e_i)),
\]
which leads to the symbol
\[
\sigma_{\Delta_{\sZ^n}}(t) := \frac{1}{2n} \sum_{i=1}^n (t_i +
t_i^{-1}), \quad t \in \sT^n.
\]
Hence, $\spec \Delta_{\sZ^n} = [-1, \, 1]$. \hfill \qed
\end{example}
\begin{example}[The zigzag graph] \label{ex1}
Let $\Gamma = (X , \sim)$ be the zigzag graph in the plane $\sR^2$
as shown in Figure \ref{fig1}. The graph $\Gamma$ is periodic with
respect to the action $g \cdot x_n := x_{n + 2g}$ of the group
$\sZ$, and the set $\cM = \{x_1, \, x_2\}$ of vertices represents
the fundamental domain.

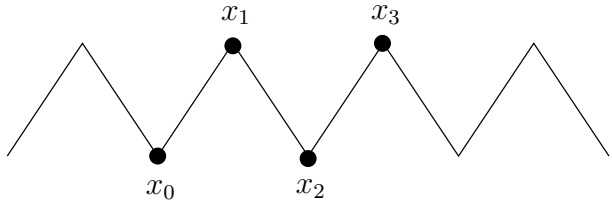
\begin{figure}[htbp] \label{fig1}
\begin{center}
\input{zickzack.pstex_t}
\end{center}
\end{figure}

One should mention that, as a graph, the zigzag graph is
isomorphic to the Cayley graph of the group $\sZ$ and, in
both cases, it is the same group $\sZ$ which acts on the graph. 
The difference lies in the way in which $\sZ$ acts. For the Cayley
graph, the group element $\alpha$ maps the vertex $x$ to $\alpha +
x$, whereas $\alpha$ maps $x$ to $2 \alpha + x$ for the zigzag
graph. The latter action is visualized by the zigzag form.

The operator $U \Delta_\Gamma U^{-1}$ has the matrix
representation 
\[
U \Delta_\Gamma U^{-1} = \frac{1}{2} \, \left(
\begin{array}{cc}
0 & I + V_{(1, \, 0)} \\
I + V_{(-1, \, 0)} & 0
\end{array}
\right)
\]
in the basis induced by $\cM$. Hence,
\[
\sigma_{\Delta_\Gamma} (t) = \frac{1}{2} \, \left(
\begin{array}{cc}
0 & 1+t \\
1+t^{-1} & 0
\end{array}
\right), \quad t\in \sT,
\]
and a straightforward calculation shows that the spectral curves
of $\Delta_\Gamma$ are
\[
\{\lambda \in \sC : \lambda = \pm \cos^2 \varphi/2, \, \varphi \in
[0, \, 2\pi]\}.
\]
Hence, the spectrum of the Laplacian $\Delta_\Gamma$ of the zigzag
graph is the interval $[-1, \, 1]$.

Next consider the Schr\"{o}dinger operator $\cH_\Gamma :=
\Delta_\Gamma + vI$ with $\sZ$-periodic potential $v$. Thus, $v$ is
completely determined by its values on $\cM$, and we write $v_1 :=
v(x_1)$ and $v_2 := v(x_2)$. Then
\[
\sigma_{\cH_\Gamma - \lambda I}(t) = \left(
\begin{array}{cc}
v_1 - \lambda & (1+t)/2 \\
(1+t^{-1})/2 & v_2 - \lambda
\end{array}
\right), \quad t \in \sT,
\]
which implies that the spectral curves of $\cH_\Gamma$ are
\[
\left\{\lambda \in \sC : \lambda = \frac{1}{2} \pm \frac{ \sqrt{(v_1 -
v_2)^2 + 4\cos^2\varphi/2}}{2(v_1 + v_2)}, \; \varphi \in [0, \, 2
\pi] \right\}.
\]
If, for example, $v_1$ and $v_2$ are real numbers with $v_1 <
v_2$, then $\spe \cH_\Gamma = \spec \cH_\Gamma$ is the union of
the disjoint intervals
\begin{eqnarray} \label{3.1'}
\left[ \frac{1}{2} - \frac{\sqrt{(v_1 -v_2)^2 + 4}}{2(v_1 + v_2)},
\, \frac{v_1}{v_1 + v_2} \right]
\bigcup \left[ \frac{v_2}{v_1 + v_2}, \, %
\frac{1}{2} + \frac{\sqrt{(v_1 - v_2)^2+4}}{2(v_1 + v_2)}\right],
\end{eqnarray}
that is, one observes a gap $(\frac{v_1}{v_1 + v_2}, \,
\frac{v_2}{v_1 +v_2})$ in the spectrum.

Finally, let the potential $v$ be slowly oscillating. Then the
essential spectrum of $\cH_\Gamma$ is the union
\begin{eqnarray} \label{3.2'}
\lefteqn{\bigcup_h \left[ \frac{1}{2} -
\frac{\sqrt{(v_1^h-v_2^h)^2 + 4}}{2(v_1^h+v_2^h)}, \,
\frac{\min \, \{v_1^h, \, v_2^h\}}{v_1^h + v_2^h}\right]} \\
&& \bigcup_h \left[\frac{\max \, \{v_1^h, \, v_2^h\}} {v_1^h +
v_2^h}, \, \frac{1}{2} +
\frac{\sqrt{(v_1^h-v_2^h)^2+4}}{2(v_1^h+v_2^h)}\right] \nonumber
\end{eqnarray}
where the unions are taken with respect to all sequences $h$ for
which the limits
\begin{equation} \label{3.3'}
v_j^h := \lim_{m \to \infty} v(h(m) \cdot x_j), \quad j = 1, \, 2,
\end{equation}
exist. Set
\[
a_{\cH_\Gamma} := \limsup_{\sZ \ni \alpha \to \infty}
\frac{v(\alpha \cdot x_1)}{v(\alpha \cdot x_1) + v(\alpha \cdot
x_2)},
\]
\[
b_{\cH_\Gamma} := \liminf_{\sZ \ni \alpha \to \infty}
\frac{v(\alpha \cdot x_2)}{v(\alpha \cdot x_1) + v(\alpha \cdot
x_2)}.
\]
Thus, if the inequality
\begin{equation} \label{3.4}
a_{\cH_\Gamma} < b_{\cH_\Gamma}
\end{equation}
holds, then the operator $\cH_\Gamma$ has the gap
$(a_{\cH_\Gamma}, \, b_{\cH_\Gamma})$ in its essential spectrum.
Of course, this interval can contain points of the discrete
spectrum of $\cH_\Gamma$. \hfill \qed
\end{example}
\begin{example}[The honeycomb graph] \label{ex2}
Let $\Gamma = (X, \, \sim)$ be the hexagonal graph shown in
Figure \ref{fig2}. We consider this graph as embedded into $\sR^2$
and let $e_1$ and $e_2$ be the vectors indicated in the figure.
The group $\sZ^2$ operates on $\Gamma$ via
\[
(\alpha_1, \alpha_2) \cdot x := x + \alpha_1 e_1 + \alpha_2 e_2
\]
(where $\alpha_1, \, \alpha_2 \in \sZ$ and $x \in X$). A
fundamental domain $\cM$ for this action is provided by any two
vertices $x_1, \, x_2$ as marked in the figure.

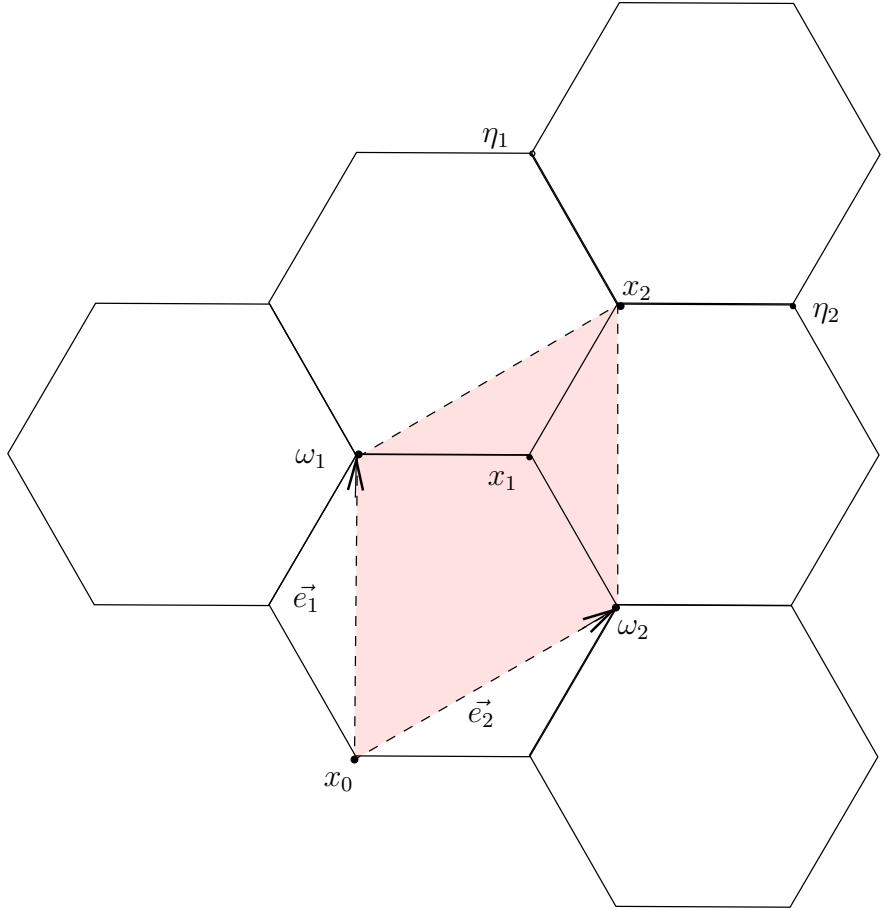
\begin{figure}[htbp] \label{fig2}
\begin{center}
\input{sechseck.pstex_t}
\end{center}
\end{figure}

Hence, we have to identify $l^p(X)$ with $l^p(\sZ^2, \, \sC^2)$,
and the Laplacian $\Delta_\Gamma$ has the following matrix
representation with respect to $\cM$
\[
U \Delta_\Gamma U^{-1} = \frac{1}{3} \, \left(
\begin{array}{cc}
0 & I + V_{e_1} + V_{e_2} \\
I + V_{e_1}^{-1} + V_{e_2}^{-1} & 0
\end{array}
\right).
\]
Consequently,
\[
\sigma_{\Delta_\Gamma} (t) = \frac{1}{3} \, \left(
\begin{array}{cc}
0 & 1 + t_1 + t_2 \\
1 + t_1^{-1} + t_2^{-1} & 0
\end{array}
\right), \quad t = (t_1, \, t_2) \in \sT^2,
\]
and the spectral curves of the Laplacian $\Delta_\Gamma$ are
\[
\cC_\pm := \{\lambda \in \sC : \lambda = \pm |1 + e^{i \varphi_1}
+ e^{i \varphi_2}|/3, \; \varphi_1, \, \varphi_2 \in [0, \, 2\pi]
\}.
\]
The curves $\cC_\pm$ coincide with the intervals $[0, \, 1]$ and
$[-1, \, 0]$, respectively, whence $\spec \Delta_\Gamma = [-1, \,
1]$.

Let now $v$ be a $\sZ^2$-periodic potential and set $v_j := 
v(x_j)$ for $j = 1, \, 2$. A calculation similar to Example 
\ref{ex1} yields that the spectral curves of the Schr\"{o}dinger 
operator $\cH_\Gamma := \Delta_\Gamma + vI$ are
\[
\left\{\lambda \in \sC : \lambda = \frac{1}{2} \pm \frac{
\sqrt{(v_1-v_2)^2 + 4\mu(\varphi_1, \, \varphi_2)}}{2(v_1+v_2)}
\right\},
\]
where
\[
\mu(\varphi_1, \, \varphi_2) := |1 + e^{i \varphi_1} + e^{i
\varphi_2}|^2/9, \quad \varphi_1, \, \varphi_2 \in [0, \, 2\pi].
\]
Hence, as in Example \ref{ex1}, $\spe \cH_\Gamma = \spec
\cH_\Gamma$ is given by the union (\ref{3.1'}). Let finally $v$ be
a slowly oscillating potential on $X$. Since the image of the
function $\mu$ is the interval $[0, \, 1]$, the essential spectrum
of the Schr\"{o}dinger operator on the honeycomb graph $\Gamma$ is
given by formulas (\ref{3.2'}) and (\ref{3.3'}). If the condition
(\ref{3.4}) holds, then a gap $(a_{\cH_\Gamma}, \,
b_{\cH_\Gamma})$ occurs in the essential spectrum of $\cH_\Gamma$.
\hfill \qed
\end{example}
\section{A three-particle problem} \label{sect7}
Let $\Gamma := (X, \, \sim)$ be a $\sZ^n$-periodic discrete graph.
We consider the Schr\"{o}dinger operator
\begin{eqnarray} \label{4.6}
\cH u & := & \Delta_\Gamma \otimes I_X + I_X \otimes \Delta_\Gamma +   \\
&& \qquad + (W_1 I_X) \otimes I_X + I_X \otimes (W_2 I_X) +
W_{12}I \nonumber
\end{eqnarray}
on $l^2 (X \times X)$. This operator describes the motion of two
particles with coordinates $x^1, \, x^2 \in X$ with masses $1$ on
the graph $\Gamma$ around a heavy nuclei located at the point 
$x_0 \in X$. Therefore, $\cH$ is also called a
3-particle Schr\"{o}dinger operator. In (\ref{4.6}),
$\Delta_\Gamma$ is again the Laplacian on the graph $\Gamma$,
$I_X$ is the identity operator on $l^2 (X)$, $I = I_X \otimes I_X$
is the identity operator on $l^2(X \times X)$, $W_1$ and $W_2$ are
real-valued functions on $X$ defined by
\[
W_j (x^j) = w_j (\rho (x^j, \,  x_0)), \quad j= 1, \, 2,
\]
and $W_{12}$ is a real-valued function on $X \times X$ given by
\[
W_{12}(x^1, \, x^2) = w_{12}(\rho (x^1, \, x^2)).
\]
Here $\rho$ denotes the given metric on $X$, and $w_1, \, w_2$ and
$w_{12}$ are functions on the real interval $[0, \, \infty)$ which
satisfy
\[
\lim_{z \to \infty} w_1 (z) = \lim_{z \to \infty} w_2 (z) =
\lim_{z \to \infty} w_{12} (z) = 0.
\]
Clearly, $\cH$ is a band operator on $l^2(X \times X)$. We are
going to describe its essential spectrum via formula (\ref{3.2}),
for which we need the limit operators of $\cH$ and their spectra.
Note that the spectrum of the Laplacian $\Delta_\Gamma$ depends on
the structure of the graph $\Gamma$ and that this spectrum has a
band structure (= is the union of closed intervals). In Examples
\ref{ex-1} -- \ref{ex2} we had $\spec \Delta_\Gamma = [-1, \, 1]$.

We agree upon the following notation. For non-empty subsets $E, \,
F$ of $\sR$, we let
\[
E+F := \{z \in \sR : z = x+y, \, x \in E, \, y \in F\}
\]
denote their algebraic sum, and we set $2E := E+E$. 

Let $g = (g^1, \, g^2) : \sN \to \sZ^n \times \sZ^n$ be a sequence
tending to infinity. We have to distinguish the following cases
(all other possible cases can be reduced to these cases by passing
to
suitable subsequences of $g$): \\[2mm]
{\bf Case 1.} The sequence $g^1$ tends to infinity, whereas $g^2$
is constant. Then the limit operator $\cH_g$ of $\cH$ is unitarily
equivalent to the operator
\begin{equation} \label{e4.7}
\cH_2 := \Delta_\Gamma \otimes I_X + I_X \otimes (\Delta_\Gamma +
W_2 I_X).
\end{equation}
{\bf Case 2.} Here $g^2$ tends to infinity and $g^1$ is constant.
Then the limit operator $\cH_g$ of $\cH$ is unitarily equivalent
to the operator
\begin{equation} \label{e4.8}
\cH_1 := (\Delta_\Gamma + W_1 I_X) \otimes I_X + I_X \otimes
\Delta_\Gamma.
\end{equation}
{\bf Case 3.} Both $g^1$ and $g^2$ tend to infinity. There
are two subcases: \\[1mm]
{\bf Case 3a.} The sequence $g^1 - g^2$ tends to infinity. In this
case the limit operator is the free discrete Hamiltonian
\[
\Delta_\Gamma \otimes I_X + I_X \otimes \Delta_\Gamma
\]
the spectrum of which is equal to $2 \, \spec \Delta_\Gamma$. \\[1mm]
{\bf Case 3b.} The sequence $g^1 - g^2$ is constant. Then the
limit operator $\cH_g$ of $\cH$ is unitarily equivalent to the
operator of interaction
\begin{equation} \label{4.10}
\cH_{12} := \Delta_\Gamma \otimes I_X + I_X \otimes \Delta_\Gamma
+ W_{12}I.
\end{equation}
Note that the operators $\cH_1, \, \cH_2$ and $\cH_{12}$ are
invariant with respect to shifts by elements of the form $(0, \,
g), \, (g, \, 0)$ and $(g, \, g)$ of $\sZ^n \times \sZ^n$,
respectively. It follows from Proposition \ref{p1.4} that these
operators do not possess discrete spectra. From formula
(\ref{3.2}) we further conclude
\begin{equation} \label{4.11}
\spe \cH = \spec \cH_1 \cup \spec \cH_2 \cup \spec \cH_{12}.
\end{equation}
The following proposition is well known. For a proof see \cite{SR1},
Theorem VIII.33 and its corollary.
\begin{proposition}
Let $A \in \cL(H)$ and $B \in \cL(K)$ be bounded self-adjoint 
operators on Hilbert spaces $H, \, K$. Then
\[
\spec (A \otimes I_K + I_H \otimes B) = \spec A + \spec B.
\]
\end{proposition}
This proposition implies in our setting that
\[
\spec \cH_2 = \spec \Delta_\Gamma + \spec (\Delta_\Gamma + W_2
I_X).
\]
Since the Schr\"{o}dinger operator $\Delta_\Gamma + W_2 I_X$ is a
compact perturbation of the Laplacian $\Delta_\Gamma$, one has
\[
\spe (\Delta_\Gamma + W_2 I_X) = \spec \Delta_\Gamma  \cup
\{\lambda_k^{(2)}\}_{k=1}^\infty
\]
where $\{ \lambda_k^{(2)}\}_{k=1}^\infty$ is the sequence of the
eigenvalues of $\Delta_\Gamma + W_2 I_X$ which are located outside
the spectrum of $\Delta_\Gamma$. Thus,
\[
\spec \cH_2 = 2 \, \spec \Delta_\Gamma + \cup_{k=1}^\infty
(\lambda_k^{(2)} + \spec \Delta_\Gamma).
\]
In the same way one finds
\[
\spec \cH_1 = 2 \, \spec \Delta_\Gamma + \cup_{k=1}^\infty
(\lambda_k^{(1)} + \spec \Delta_\Gamma)
\]
where the $\lambda_k^{(1)}$ run through the points of the discrete
spectrum of $\Delta_\Gamma + W_1 I_X$ which are located outside
the spectrum of $\Delta_\Gamma$.

Recall that in Examples \ref{ex-1} -- \ref{ex2}, $\spec
\Delta_\Gamma = [-1, \, 1]$. Hence, in the context of these
examples,
\[
\spec \cH_j = [-2, \, 2] \bigcup_{k=1}^\infty [\lambda_k^{(j)} -
1, \, \lambda_k^{(j)} + 1].
\]
One can also give a simple estimate for the location of the
spectrum of $\cH_{12}$ by means of the following well-known result
(see, e.g., \cite{Lax}, p. 357).
\begin{proposition} \label{Pr3.1}
Let $A$ be a bounded self-adjoint operator on the Hilbert space
$H$. Then $\{a, \, b\} \subseteq \spec A \subseteq [a, \, b]$ where
\[
a := \inf_{\|h\| = 1} \langle Ah, \, h \rangle,            \quad
b: = \sup_{\|h\| = 1} \langle Ah, \, h \rangle.
\]
\end{proposition}
This observation implies the following inclusions for the spectra
of the operators $\cH_1, \, \cH_2$ and $\cH_{12}$. For $j = 1, \,
2$ one has
\[
2 \, \spec \Delta_\Gamma \subseteq \spec \cH_j \subseteq 2 \,
\spec \Delta_\Gamma + \left[ \inf_{x \in X} W_j (x), \, \sup_{x \in X}
W_j (x) \right],
\]
whereas
\[
2 \, \spec \Delta_\Gamma \subseteq \spec \cH_{12} \subseteq 2 \,
\spec \Delta_\Gamma + \left[ \inf_{y \in X \times X} W_{12}(y), \,
\sup_{y \in X \times X} W_{12}(y) \right].
\]
In the context of Examples \ref{ex-1} -- \ref{ex2}, these
inclusions specify to
\[
[-2, \, 2] \subseteq \spec \cH_j \subseteq \left[ -2 + \inf_{x \in X}
W_j (x), \, 2 + \sup_{x \in X} W_j (x) \right],
\]
\[
[-2, \, 2] \subseteq \spec \cH_{12} \subseteq \left[ -2 + \inf_{x \in X
\times X} W_{12} (x), \, 2 + \sup_{x \in X \times X} W_{12} (x) \right].
\]
Thus, Theorem \ref{t3.1} yields for these examples
\[
\spe \cH \subseteq [m-2, \, M+2]
\]
where
\[
m := \min \, \left\{\inf_{x \in X} W_1(x), \, \inf_{x \in X} W_2(x), \,
\inf_{x \in X \times X} W_{12}(x) \right\},
\]
\[
M := \max \, \left\{\sup_{x \in X} W_1(x), \, \sup_{x \in X} W_2(x), \,
\sup_{x \in X \times X} W_{12}(x) \right\}.
\]

\begin{thebibliography}{99}
\bibitem{AlbLakMum}
S. Albeverio, S. N. Lakaev, Z. I. Muminov, {\em On the structure of
the essential spectrum of the three-particle Schr\"{o}dinger
operators on a lattice}. arXiv:math-ph/0312050,
2003.
\bibitem{AlbaverioLakaev}
S. Albeverio, S. N. Lakaev, J. I. Abdullaev, {\em On the
finiteness of the discrete spectrum of four-particle lattice
Schrodinger operators}. Reports Math. Phys. {\bf 51}(2003), 1, 43
- 70.
\bibitem{AMP}
W. Amrein, M. M\u{a}ntoiu, R. Purice, {\em Propagation properties
for Schr\"{o}dinger operators affiliated with certain
$C^*$-algebras}. Ann. H. Poincar\'{e} In-t {\bf 3}(2002), 6, 1215
-- 1232.
\bibitem{Berkol}
G. Berkolaiko, R. Carlson, S. Fulling, P. Kuchment (Editors), {\em
Quantum Graphs and Their Applications}. Contemp. Math. {\bf 415},
Amer. Math. Soc., Providence, R.I., 2006.
\bibitem{Cycon}
H. L. Cycon, R. G. Froese, W. Kirsch, B. Simon, {\em
Schr\"{o}dinger Operators with Applications to Quantum Mechanics
and Global Geometry}. Springer-Verlag, Berlin, Heidelberg, New
York 1987.
\bibitem{Deift}
P. Deift, {\em Orthogonal Polynomials and Random Matrices: A
Rieman-Hilbert Approach}. Courant Lectures Notes Math. {\bf 3},
Amer. Math. Soc., Providence, R.I., 2000.
\bibitem{Gorgescu2}
V. Georgescu, A. Iftimovici, {\em Crossed products of
$C^*$-algebras and spectral analysis of quantum Hamiltonians}.
Comm. Math. Phys. {\bf 228}(2002), 519 -- 560.
\bibitem{Gorgescu1}
V. Georgescu, A. Iftimovici, {\em Localization at infinity and
essential spectrum of quantum Hamiltonians}.
arXiv:math-ph/0506051v1, 20 June 2005.
\bibitem{Harris}
P. Harris, {\em Carbon Nano-tubes and Related Structure}.
Cambridge Univ. Press, Cambridge 2002.
\bibitem{Jirari}
A. Jirari, {\em Second Order Sturm-Liouville Difference Equations
and Orthogonal Polynoms}. Memoirs Amer. Math. Soc. {\bf 542},
Amer. Math. Soc., Providence, R.I., 1995.
\bibitem{Kato}
T. Kato, {\em Perturbation Theory for Linear Operators}.
Springer-Verlag, Berlin, Heidelberg, New York 1966.
\bibitem{Kor1}
E. Korotyaev, I. Lobanov, {\em Schr\"{o}dinger operators on zigzag
graphs}. arXiv: math.SP/06040006.
\bibitem{Kor2}
E. Korotyaev, I. Lobanov, {\em Zigzag periodic nanotube in
magnetic field}, arXiv: math.SP/06040007.
\bibitem{KuchSp}
P. Kuchment, {\em On some spectral problems of mathematical
physics}. In: Partial Differential Equations and Inverse Problems,
C. Conca, R. Manasevich, G. Uhlmann, M. S. Vogelius (Editors),
Contemp. Math. {\bf 362}, Amer. Math. Soc., Providence, R.I.,
2004.
\bibitem{Kuch1}
P. Kuchment (Editor), {\em Quantum graphs and their applications}.
Special issue of Waves in Random Media {\bf 14}(2004), no. 1.
\bibitem{Kuch2}
P. Kuchment, {\em Quantum graphs I. Some basic structure}. Waves
in Random Media {\bf 14}(2004), 107 -- 128.
\bibitem{Kuch3}
P. Kuchment, {\em Quantum graphs II. Some spectral properties of
quantum and combinatorial graphs}. J. Phys. A {\bf 38}(2005), 22,
4887 -- 4900.
\bibitem{KuchPost}
P. Kuchment, O. Post, {\em On the spectra of carbon
nano-structures}. arXiv: math-ph/0612021v4, 19 Jan 2007.
\bibitem{KuchVainberg}
P. Kuchment, B. Vainberg, {\em On the structure of eigenfunctions
corresponding to embedded eigenvalues of locally perturbed
periodic graph operators}. Comm. Math. Phys. {\bf 268}(2006), 673
-- 686.
\bibitem{LakaevMuminov}
S. N. Lakaev, Z. I. Muminov, {\em The asymptotics of the number of
eigenvalues of a three-particle lattice Schr \"{o}dinger
operator}. Func. Anal. Appl. {\bf 37}(2003), 3, 228 -- 231.
\bibitem{LastSimon}
Y. Last, B. Simon, {\em The essential spectrum of Schr\"{o}dinger,
Jacobi, and CMV operators} (2005). Preprint {\bf 304} at \\
http://www.math.caltech.edu/people/biblio.html.
\bibitem{Lax}
P. D. Lax, {\em Functional Analysis}. Wiley-Interscience 2002.
\bibitem{Mantoiu}
M. M\u{a}ntoiu, {\em $C^*$-algebras, dynamical systems at infinity
and the essential spectrum of generalized Schr\"{o}dinger
operators}. J. Reine Angew. Math. {\bf 550}(2002), 211 -- 229.
\bibitem{Mattis}
D. C. Mattis, {\em The few-body problem on a lattice}. Rev. Modern
Phys. {\bf 58}(1986), 361 -- 379.
\bibitem{Mogilner}
A. Mogilner, {\em Hamiltonians in solid state physics as
multiparticle discrete Schr\"{o}dinger operators: Problems and
results}. Adv. Soviet Math. {\bf 5}, Amer. Math. Soc., Providence,
R.I., 1991.
\bibitem{Paul}
L. Pauling, {\em The diamagnetic anisotropy of aromatic
moleculas}, J. Chem. Phys. {\bf 4}(1936), 673 -- 677.
\bibitem{RJMP}
V. S. Rabinovich, {\em Essential spectrum of perturbed
pseudodifferential operators. Applications to the Schr\"{o}dinger,
Klein-Gordon, and Dirac operators}. Russian J. Math. Phys. {\bf
12}(2005), 1, 62 -- 80.
\bibitem{RRActa}
V. S. Rabinovich, S. Roch, {\em Pseudodifference operators on
weighted spaces and applications to discrete Schr\"{o}dinger
operators}. Acta Appl. Math. {\bf 84}(2004), 55 -- 96.
\bibitem{RRJP}
V. S. Rabinovich, S. Roch, {\em The essential spectrum of
Schr\"{o}dinger operators on lattices}. J. Phys. A: Math. Gen.
{\bf 39}(2006), 8377 -- 8394.
\bibitem{RRR1}
V. S. Rabinovich, S. Roch, J. Roe, {\em Fredholm indices of
band-dominated operators}. Integral Eq. Oper. Theory {\bf 49}(2004), 
2, 221 -- 238.
\bibitem{RRS1}
V. S. Rabinovich, S. Roch, B. Silbermann, {\em Fredholm theory and
finite section method for band-dominated operators}. Integral
Equations Oper. Theory {\bf 30}(1998), 452 -- 495.
\bibitem{RRS2}
V. S. Rabinovich, S. Roch, B. Silbermann, {\em Band-dominated
operators with operator-valued coefficients, their Fredholm
properties and finite sections}. Integral Eq. Oper. Theory {\bf
40}(2001), 3, 342 -- 381.
\bibitem{RRSB}
V. S. Rabinovich, S. Roch, B. Silbermann, {\em Limit Operators and
their Applications in Operator Theory}, Oper. Theory: Adv. Appl.
{\bf 150}, Birkh\"{a}user, Basel 2004.
\bibitem{SR1}
M. Reed, B. Simon, {\em Methods of Modern Mathematical Physics
I: Functional Analysis}, Academic Press, New York 1972.
\bibitem{SR3}
M. Reed, B. Simon, {\em Methods of Modern Mathematical Physics
III: Scattering Theory}, Academic Press, New York 1979.
\bibitem{Roch1}
S. Roch, {\em Band-dominated operators on $l^p$-spaces: Fredholm
indices and finite sections}. Acta Sci. Math. (Szeged) {\bf
70}(2004), 783 -- 797.
\bibitem{RuSch}
K. Ruedenberg, C. W. Scherr, {\em Free-electron network model for
conjugated systems I. Theory}. J. Chem. Phys. {\bf 21}(1953), 9,
1565 -- 1581.
\bibitem{Saito}
R. Saito, G. Dresselhaus, M. S. Dresselhaus, {\em Physical
Properties of Carbon Nanotubes}. Imperial College Press, London
1998.
\bibitem{Shubindis}
M. A. Shubin, {\em Discrete magnetic Laplacian}. Comm. Math. Phys.
{\bf 164} (1994), 2, 259 -- 275.
\bibitem{Yafaev}
D. R. Yafaev, {\em Scattering Theory: Some Old and New Problems}.
Lecture Notes Math. {\bf 1735}, Springer-Verlag, Berlin 2000.
\end{thebibliography}
\end{document}